\documentclass{article}
\usepackage{fullpage}

\usepackage[tight]{units}
\pdfoutput=1
\usepackage{arxiv}
\usepackage[utf8]{inputenc} 
\usepackage[T1]{fontenc}    
\usepackage{hyperref}       
\usepackage{url}            
\usepackage{booktabs}       
\usepackage{amsfonts}       
\usepackage{nicefrac}       
\usepackage{microtype}      
\usepackage{lipsum}
\usepackage{graphics,graphicx}
\usepackage{pgfplots}
\usepackage{caption}
\usepackage{subfig}
\usepackage{tikz}
\usepackage{ifpdf}
\usepackage{amsmath, calc, xcolor}
\usepackage{mathtools}
\usepackage{pgf}

\newlength\dlf
\newcommand\alignedbox[2]{
	&
	\begingroup
	\settowidth\dlf{$\displaystyle #1$}
	\addtolength\dlf{\fboxsep+\fboxrule}
	\hspace{-\dlf}
	\fcolorbox{black}{gray!15}{$\displaystyle #1 #2$}
	\endgroup
}

\pgfplotsset{width=8cm,compat=1.8}
\title{An approach of the interior of a black hole type Schwarzschild}

\author{
	J. Enrique H. Ramírez \\
	Department of Physics\\
	ABC Federal University\\
	Sao Paulo, Brazil\\
	\texttt{enrique.ramirez@ufabc.edu.br} \\
}

\begin{document}
	\maketitle
	
	\begin{abstract}
		Our goal in this paper is to explain the internal properties of the black hole by considering its density as a function of the reciprocal of its radius and the temperature as a function of the reciprocal of the density. We then set the temperature to the Hawking temperature. This gives all the macroscopic quantities of the black hole, such as heat capacity, pressure, surface gravity, and equation of state. In this work, we only consider a black hole with mass $M$, radius $r_{+}$ and vacuum. Two internal forces and the corresponding potential are obtained. In addition, the potential of the vacuum, the potential of the black hole, the effective potential, the density of the vacuum and the effective density are obtained. A possible nexus of the model developed in this work with the observational data is also shown. The rotation curve of the galaxies around the clusters is shown, which fits the experimental curve.
	\end{abstract}
	
	
	\section{Introduction}
	According to the general theory of relativity, black holes have no internal structure. Its characteristics are established from the geometric properties of space-time that defines the black hole, specifically its event horizon. All the analysis we do in this work is under the assumption that the black hole has an internal structure. Recent observations have been able to show indirect images of a black hole, but this remains a mystery and even more so its internal structure. In this sense, if we can't observe it directly and know its internal structure and properties, then we must imagine it. Experience shows that the mass density is by no means a constant quantity, in general it is not homogeneous or isotropic. Under certain conditions we can assume that objects have a constant mass density, but for large objects, such as planets or stars, this fact is not true, the density increases in the depth of the large objects. For this reason, we consider density to be proportional to $1/r^n$, where $n$ is a specific integer yet to be determined. On the other hand, as we know, the temperature is a consequence of the motion of the atoms or particles, that is, the temperature has to do with the kinetic energy of the atoms or particles. If the freedom of the particles or atoms to move in the container is reduced, we hope that the temperature will decrease. For this reason, we consider that the temperature is the inverse of the density.
	
	Dimensional analysis is used to obtain the density, temperature and entropy of the black hole. We obtain the surface gravity, the head capacity, the temperature as a function of area, two forces that determine the pressure and motion in the black hole, and the potential energy associated with these forces. The black hole potential as well as the vacuum potential and the effective potential are obtained. The energy density of the vacuum is determined, as well as the effective density. A link is found between the energy density of the vacuum and the energy density of the object of analysis(the black hole).This relationship allows us to explain the principle of cosmic coincidence and answer the question of why, despite the universe expanding, the orders of magnitude of the energy density of a vacuum and ordinary matter remain the same. It is known that repulsive properties are attributed to the energy density of a vacuum. Surprisingly in this work we discovered that this density also shows attractive properties in a certain region, which we associate with the possible origin of dark matter. This new property allows us to determine the attractive contribution of mass from the energy density of the vacuum, which helps to explain the missing and non-visible mass found in the clusters, which allows us, in turn, to correctly reproduce the rotation curve of the galaxies. The percentages of the referred densities are determined and it is found that they correspond to the percentages reported in the literature ($75\%$ for dark energy and $25\%$ for matter with attractive properties).In the latter case, we were also able to determine the percentage that corresponds to baryonic matter ($6\%$) and that which corresponds to the attractive contribution from the vacuum energy density ($19\%$), which we identify as dark matter. Although the model was designed to try to characterize the interior of a black hole, surprisingly it finds application in another area such as cosmology. This link is established by means of the energy density of the vacuum with the cosmological constant and with the Hubble constant. 
	
	Using observational data we were able to determine, using the model proposed in this work, the energy density of the vacuum, with a value very close to those reported, the cosmological constant and the Hubble constant, also with values very close to those reported in the literature. The model finds a functional dependence on the Hubble constant, which makes it no longer constant. On the other hand, it succeeds in reproducing the law with which the galaxies move away, showing that for certain regions this dependence is linear in appearance. In addition, the rotation curve of the galaxies is reproduced.
	
	\section{Mass density}
	We assume that the density of the black hole is proportional to $ 1/ r^{n}$, i. e:
	
	\begin{equation}\label{density}
	\rho = \frac{a k}{r^{n}},
	\end{equation}
	
	where $k$ is dimensionless, $n$ is a positive integer, and $a$ is a certain function $a = a(G,c)$, where $G$ and $c$ are the gravitational and speed-of-light constants, respectively.\
	
	Let $a$ be
	
	\begin{equation}\label{aproportional}
	a = G^{x} c^{y},
	\end{equation}
	
	then, through a simple dimensional analysis we obtain the exponents $x$, $y$ and $n$, where $x = -1$, $y = 2$, and $n = 2$, then 
	
	\begin{equation}\label{densityexpresion}
	\rho = k \frac{c^{2}}{G r^2}.
	\end{equation}
	
	Let $M$ and $r_{+}$ be the mass and the Schwarzschild radius of the black hole. The integral of Eq. (\ref{densityexpresion}) let us determine the $k$ constant.
	
	\begin{equation}
	M = k \int_{2 l_{P}}^{r_{+}} \frac{c^{2}}{G r^2} 4 \pi r^{2} dr
	\end{equation}
	
	where $l_{P}$ is the Planck length and  $r_{+} = 2 G M/c^{2}$.
	
	Then we find,
	
	\begin{equation}
	k = \frac{1}{8 \pi},
	\end{equation}
	
	so
	
	\begin{equation}\label{densityfinal}
	\rho = \frac{1}{8 \pi} \frac{c^{2}}{G r^2},
	\end{equation}
	
	where $2 l_{P} \leq r < \infty $.
	
	$\rho$ is the volumetric mass density, but can be seen as the surface mass density if we take Eq. (\ref{densityfinal}) as
	
	\begin{equation}
	\rho = \frac{M}{4 \pi r^{2}}\frac{1}{r_{+}}
	\end{equation}
	In this way the mass $m$ enclosed by volume $4/3 \,\pi r^3$ is given by
	
	\begin{equation}\label{massinsideofblackhole2}
	m = \frac{c^2}{2 G}r.
	\end{equation}
	
	This would be the mass that would correspond to each black hole in the formation process from $2 l_{P}$(whose corresponding mass is the Planck mass, $m_{P}$) to $r_{+}$.
	
	\section{Temperture}
	In a similar way as was done in the previous section, wwe now consider that the temperature depends on the inverse of the density, where  $a = a[G,k_{B},\hbar,c]$ is the coefficient of proportionality,  being $k_{B}$ and $\hbar$ the Bolztmann and Planck constants respectively.
	
	\begin{equation}\label{temperaruredenditydefinicion}
	T = c_{0}\frac{a}{\rho},
	\end{equation}
	
	where $c_{0}$ is a constant that has no dimension, then,
	
	\begin{equation}\label{temperaruredendity}
	T = c_{0}\frac{8 \pi a G}{c^2}r^2.
	\end{equation}
	
	In the dimensional analysis, we find the unknowns $x,y,z$ and $w$ from the dimensional relation and we get $x =-5/2$, $y = -1$, $z = -1/2$, and $w = 15/2$, that is, $a = G^{-5/2} k_{B}^{-1} \hbar^{-1/2} c^{15/2}$.
	
	Substituting these values into Eq. (\ref{temperaruredendity}), one obtains,
	
	\begin{equation}\label{tempcuasifinal}
	T = c_{0} \frac{8 \pi c^{11/2}}{k_{B} \hbar^{1/2} G^{3/2}}r^{2}.
	\end{equation}
	
	To find the $c_{0}$ constant, we consider that $T[r_{+}] = T_{H}$, where $T_{H}$ is the Hawwking temperature of the black hole \cite{hawking1976black}. Then we find that,
	
	\begin{eqnarray}
	c_{0} = \frac{1}{256 \pi^2}\Big[\frac{(\frac{\hbar c}{G})^{1/2}}{M}\Big]^3.\nonumber
	\end{eqnarray}
	
	Substituting this expression into Eq. (\ref{tempcuasifinal}) leads to,
	
	\begin{eqnarray}\label{temperatureenrique}
	T =  \frac{\hbar c}{4 \pi k_{B} r_{+}^{3}}r^{2}.
	\end{eqnarray}
	
	For a given black hole of mass $m$, this temperature, Eq. (\ref{temperatureenrique}) , increases as $r$ does.The maximum value of $T$ occurs at $r = r_{+}$, the Hawking temperature $T_{H}$.If we assume that the black hole has internal structure, then this temperature, Eq. (\ref{temperatureenrique}), states that the entire black hole is not in thermal equilibrium, each surface of area $A = 4 \pi r^2$ has its own temperature $T$, this causes continuous heat transport as the inner surfaces become cooler with decreasing $r$ and tend towards zero Kelvin which cannot be reached. The minimum temperature, from the classical point of view, inside the black hole is reached when $r = 2 l_{P}$. For a typical black hole of mass $M = 3 M_{S}$,  the temperature of its surface would be $2.0552\times 10^{-8} \unit{K}$.  Thermal equilibrium is only reached on any surface of radius $r$.
	
	\section{Entropy}
	
	The next important physical quantity of the black hole is entropy. Experience tells us that if the system is orderly, the entropy is low. If the physical system has a high density, then it is ordered, to this the entropy is proportional to the reciprocal of the density, but not only to the mass. Density plays a fundamental role in this analysis. Then,
	
	\begin{equation}\label{enropydefinition1}
	S = b\frac{c_{1}}{\rho}k_{B}, 
	\end{equation}
	
	where $b$ is a number and $c_{1}$ is a certain function of $G,\hbar$ and $c$, so $c_{1} = c_{1}(G,\hbar,c)$. Since the unit of entropy is $[S] = [k_{B}]$, then $[c_{1}] = [\rho]$.
	
	In a similar way as before, we consider a dimensional analysis, then,
	
	\begin{equation}
	[G]^x [\hbar]^y [c]^z = [\rho].
	\end{equation}
	
	Solving the resulting system, we have $x = -2$, $y = -1$ and $z = 5$. Then it follows from Eqs. (\ref{enropydefinition1}) and (\ref{densityfinal}) that,
	
	\begin{equation}\label{entropiaenrique}
	S = a \frac{8 \pi k_{B} c^3}{\hbar G}r^2.
	\end{equation}
	
	Now we have to analyze the coefficient $a$. We are dealing with an isolated system, a black hole. Except $a$, all terms occurring in Eq. (\ref{entropiaenrique}) are positive. The entropy of an isolated system increases, so $a$ must also be positive.
	
	Since $2 l_{P} < r \leq r_{+}$, the maximum value of $S$ occurs at the event horizon. If $r = r_{+}$, we have then,
	
	\begin{equation}\label{entropiaarea}
	S = a \frac{8 k_{B} c^3}{4\hbar G}A,
	\end{equation}
	
	that is the entropy of the black hole surface, which differs from the Hawking entropy, $S_{H}$, by a factor $8 a$, which coefficient $a$ will be determined in the next section.
	
	Eq.(\ref{entropiaarea}) can be interpreted as the entropy of each of the surfaces of black holes in their formation process from $r = 2 l_{P}$ to $r = r_{+}$ or as the entropy of each of the interior surfaces of the black hole in the hypothesis that it has an internal structure.
	
	\section{Heat Capacity}
	
	Imagine a process, by means of which we add a certain amount $dm$ of matter to a black hole, with which we vary both its area and its temperature. The heat capacity in any process is defined as:
	
	\begin{eqnarray}\label{capacidadcalorificadef}
	C = \frac{\delta Q}{d T}.
	\end{eqnarray}
	
	For a black hole,
	
	\begin{eqnarray}
	d \, m c^2 = \frac{c^2 \kappa}{8 \pi G}d \, A + \Omega d\, J + \Phi d \, \mathcal{Q}, 
	\end{eqnarray}
	
	where  $\kappa$, $\Omega$ , $J$ , $ \Phi$ and $\mathcal{Q}$ refer respectively to the surface gravity, its angular velocity, angular moment, electric potential and electric charge. In this work we consider $J = 0$ and $\mathcal{Q} = 0$. In \cite{hawking1973} and \cite{hawking1975particle}, Hawking showed that the term $(c^2 \kappa/{8 \pi G})d A$ is identified as $\delta Q$, so,
	
	\begin{equation}\label{capacidacalorificadef}
	C = \frac{dU}{dT}.
	\end{equation} 
	
	The internal energy of the black hole, $U$, is given by,
	
	\begin{equation}\label{Energiainternablackhole}
	U = m c^2,
	\end{equation}
	
	where $m$ is given by Eq. (\ref{massinsideofblackhole2}) and $T_{H}$ is given by :
	
	\begin{eqnarray}\label{temperatureenriquefrontera}
	T_{H} = \frac{\hbar c}{4 \pi k_{B} r_{+}}.
	\end{eqnarray}
	
	Then,
	
	\begin{equation}\label{capacidacalorifica1}
	C = \frac{dU}{dr_{+}}{\frac{dr_{+}}{dT}},
	\end{equation}
	
	and we obtain,
	
	\begin{equation}\label{capacidaduno}
	C = -\frac{m c^2}{T_{H}},
	\end{equation}
	
	showing that $C < 0$, which match in sign with \cite{ma2015stability}.
	
	On the other hand, $\delta Q = S dT$, then we can rewrite Eq. (\ref{capacidacalorificadef}) as,
	
	\begin{equation}\label{capacidadtres}
	C = T\frac{d S}{d T}.
	\end{equation}
	
	Considering the Eqs. (\ref{entropiaenrique}) and (\ref{temperatureenriquefrontera}) and that we can rewrite (\ref{capacidadtres}) as, 
	
	\begin{equation}
	C = T \frac{d S}{ d r_{+}}{\frac{d r_{+}}{d T}},
	\end{equation}
	
	we get,
	
	\begin{equation}\label{capacidaddos}
	C = -\frac{8 a m c^2}{T_{H}}.
	\end{equation}
	
	From this analysis we can see that the comparison of these two results, Eqs. (\ref{capacidacalorifica1}) and (\ref{capacidaddos}) yield that $a = 1/8$, as we hope. Then, for any $r$, taking into account (\ref{massinsideofblackhole2}) and (\ref{temperatureenriquefrontera}), we have that, 
	
	\begin{eqnarray}
	C = -\frac{ 2 \pi k_{B} c^3}{\hbar G}r^2
	\end{eqnarray}
	
\begin{figure}[h]
		\centering
		\includegraphics[width=0.6\linewidth, height=0.3\textheight]{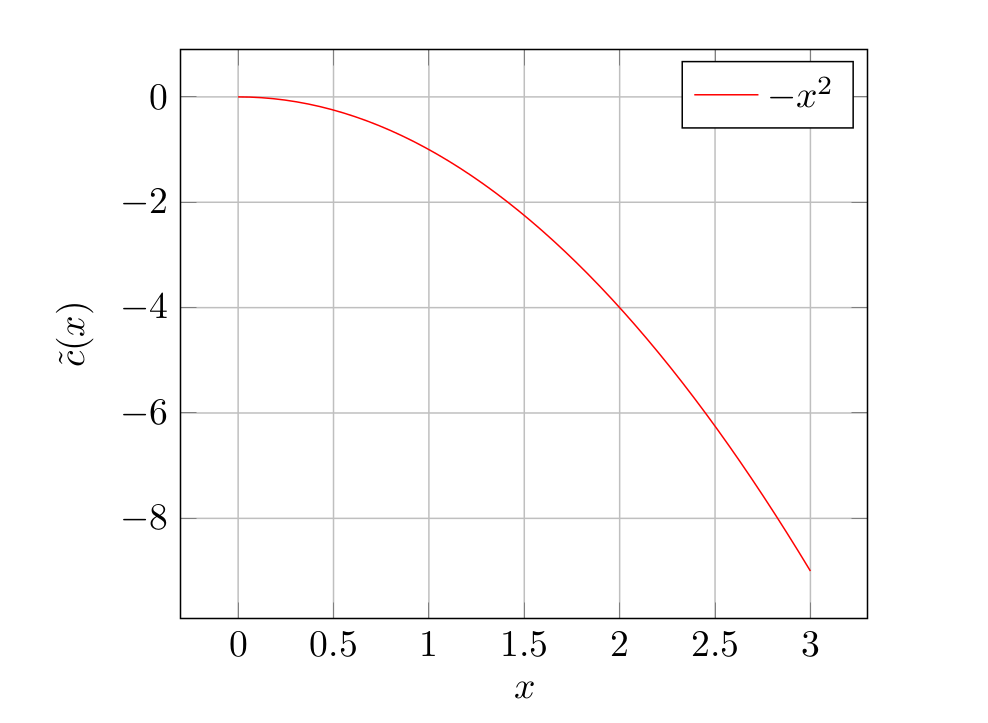}
		\caption{The heat capacity according to the model, $l_{P}/r_{+} < x  \leq 1$ and $x \equiv r/r_{+}$.}
		\label{FIGURA1}
	\end{figure}

	Fig. \ref{FIGURA1} whows this heat capacity, where $C_{0} $ is given by,
	
	\begin{eqnarray}
	C_{0} = \frac{8 \pi k_{B} G M^2}{ \hbar c}\nonumber.
	\end{eqnarray}
	
	Taking into account that $a = 1/8$, Eq. (\ref{entropiaarea}) can be written as, 
	
	\begin{equation}\label{entropiaenrique2}
	S = \frac{k_{B} c^3}{\hbar G} \pi r^2,
	\end{equation}
	
	a desired result, the Hawking entropy for the black hole.
	
	Thus, theoretically and from the classical point of view, the minimum value of the black hole entropy is given by($r = 2l_{P}$),
	
	\begin{eqnarray}
	S = 4 \pi k_{B},
	\end{eqnarray}
	
	and its maximum value by the Eq. (\ref{entropiaenrique2}) , for $r = r_{+}$.

	\section {Equation of State}
	The black hole equation of state is analysed in \cite{dolan2010cosmological}, in terms of volume and pressure. In this work, the pressure $P$ will be considered as a function of $\mathcal{A}$, since thermodynamic equilibrium exists only on the surface of the black hole. Equation of state, temperature and entropy are related to $\mathcal{A}$.
	
	We first consider the first law of thermodynamics,
	
	\begin{equation}\label{primeraleytermodynamic1}
	U = T S - P V.
	\end{equation}
	
	Solving for $P$ and considering the Eqs. (\ref{temperatureenrique}), -(\ref{Energiainternablackhole}), -(\ref{entropiaenrique2}) , $V = (1/6\sqrt{\pi} )\mathcal{A}^{3/2}$ and $r = (1/2)(\mathcal{A}/\pi)^{1/2}$ we have
	
	\begin{eqnarray}\label{presionfinal}
	P = \frac{3 F_{P}}{2 \pi }\Bigg[\frac{\Big(\frac{\mathcal{A}}{\pi}\Big)^{1/2}}{16 r_{+}^3} - \Big(\frac{\mathcal{A}}{\pi}\Big)^{-1}\Bigg],
	\end{eqnarray}
	
	where $F_{P}$ is the Planck force.
	
	If $\mathcal{A}_{+} = 4 \pi r_{+}^2$ (the event horizon),
	
	\begin{equation}\label{pressurenoevenhorizon2}
	P_{eh} = -\frac{3 P_{P}}{64 \pi}\left(\frac{m_{P}}{M}\right)^2,
	\end{equation}
	
	that is, at the event horizon, the pressure $P$ is negative, where $P_{P}$ is the Planck pressure.
	
	If $\mathcal{A} = \mathcal{A}_{min} = 16 \pi l_{P}^2$ (the minimum area from the classical point of view),
	
	\begin{eqnarray}\label{pressureareaplanck}
	P_{l_{P}} = - \frac{3 P_{P}}{64 \pi},
	\end{eqnarray}

	thus the quotient between these two pressures is of the order of $P_{lp}/P_{eh}\approx 10^{192}$, for a black hole of mass $3 M_{S}$.

	Again, assuming the black hole has internal structure, Eq. (\ref{presionfinal}) allows us to find the pressure on each surface inside a black hole as a function of the area of that surface.
	
	\newpage
	
	\section {Pressure, Force and Potential Energy}
	Again we consider Eq. (\ref{primeraleytermodynamic1}) and solve it for $P$, now taking into account that $V = 4/3\pi r^3$ and $\mathcal{A} = 4 \pi r^2$,
	
	\begin{eqnarray}\label{presiontotal}
	P = 3 c^2 \rho_{+}\left[\frac{1}{2} \left(\frac{r}{r_{+}}\right) - \left( \frac{r_{+}}{r}\right)^2\right],
	\end{eqnarray}
	
	where,
	
	\begin{align}\label{pressionblacckhole}
	\alignedbox{}{p_{bh} = \frac{3 c^2}{2}\rho_{+}\left(\frac{r}{r_{+}}\right),}
	\end{align}
	
	is the pressure due to the black hole, and

	\begin{align}\label{presionvacio}
	\alignedbox{}{p_{vac} = -3 c^2\rho_{+}\left(\frac{r_{+}}{r}\right)^2,}
	\end{align}
	
	the pressure due to the vacuum.	Then , we can rewrite \ref{presiontotal} as, 
	
	\begin{align}\label{presioncomAyr}
	\alignedbox{}{P(\mathcal{A},r) =  \frac{3 c^4}{4 r_{+}^3 G}\frac{r^3}{\mathcal{A}} - \frac{3  c ^4}{2 G}\frac{1}{\mathcal{A}}.}
	\end{align}
	
	The above expression allows us to identify two forces, one of which depends on mass and the other does not,
	\begin{eqnarray}\label{fuerzaatractiva}
	F_{bh} = \frac{3 c^4}{4  G}\Big(\frac{r}{r_{+}}\Big)^3,
	\end{eqnarray}
	
	and
	\begin{eqnarray}\label{fuerzarepulsiva}
	F_{vac} = -\frac{3 c^4}{ 2 G}.
	\end{eqnarray}
	
	The Eqs. (\ref{fuerzaatractiva}) and (\ref{fuerzarepulsiva}) are the black hole and the vacuum forces respectively, and as a result, a net force acts on the surface $\mathcal{A}$, which is given by,
	
	\begin{eqnarray}\label{fuerzanetasobresuperficie}
	F_{n} = \frac{3 c^4}{2 G}\Big[\frac{1}{2}\Big(\frac{r}{r_{+}}\Big)^3 - 1 \Big].
	\end{eqnarray}
	
	This force acts perpendicular to the surface $\mathcal{A}$ and it is radial outward from the black hole because in it $F_{vac} > F_{bh}$, then,
	
	\begin{eqnarray}\label{forzatotalvector}
	\textbf{F}_{n} = \frac{3 c^4}{2  G}\Big[\frac{1}{2}\Big(\frac{r}{r_{+}}\Big)^3 - 1 \Big]\textbf{r}.
	\end{eqnarray}
	
	On the event horizon this force is given by,
	
	\begin{eqnarray}\label{fuerzanohorizonte}
	\textbf{F}_{n}(r_{+}) = -\frac{3 c^4}{4 G}\textbf{r},
	\end{eqnarray}
	
	outward from the black hole.
	
	\begin{figure}[h]
		\centering
		\includegraphics[width=0.6\linewidth, height=0.3\textheight]{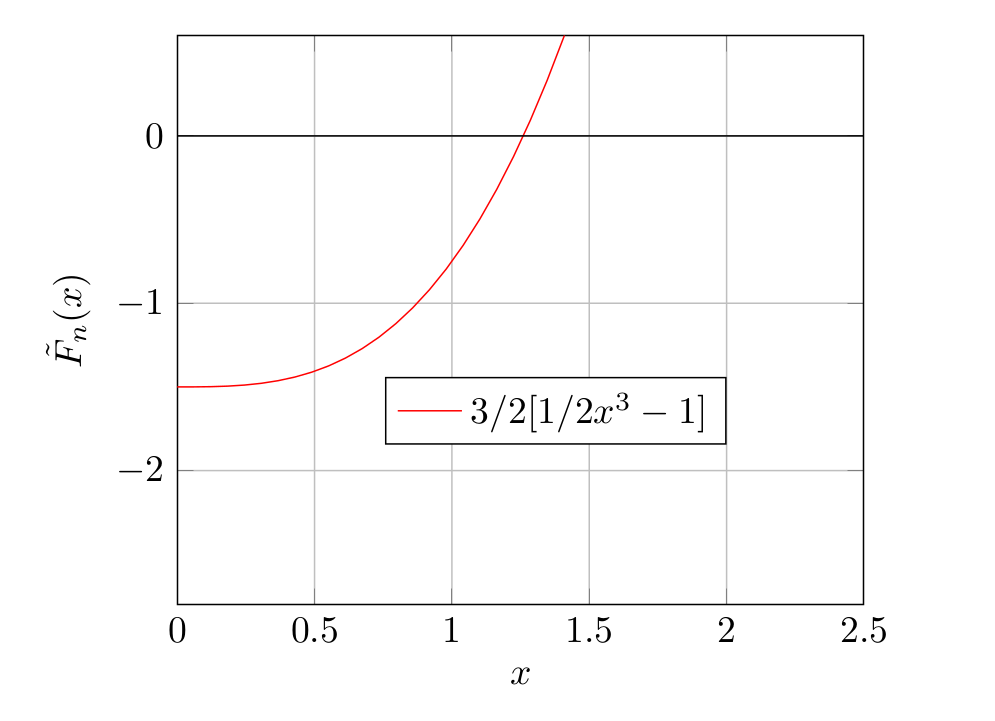}
		\caption{The net force $\tilde{F}_{n}(x)$ acting on the surface of the black hole, where $\tilde{F}_{n}(x) = F_{n}(x)/F_{0}$ and $F_{0} = c^4/G$.} \label{FIGURA2}
	\end{figure}

	Now it is clear why the pressure $P$ at the event horizon is negative. No matter how big the black hole is, this force has the same value at the event horizon. This fact has an important consequence. If we assume the entire universe to be a black hole, in the sense that no matter can escape from it, then this force tends to expand it at the event horizon. The black hole force, $F_{bh}$, depends on the mass and corresponds to an attractive force and the vacuum force, $F_{vac}$, to an attractive one in the opposite direction. 
	
	$\textbf{F}_{n}$ is of the form $\textbf{F}_{n} = F(r)\textbf{r}$ and as we know, this type of force is conservative. This means that we can derive a corresponding potential energy $U(r)$. The integral of Eq. (\ref{forzatotalvector}) from $r_{+}$ to a point $r$, inside the black hole, we obtain
	
	\begin{eqnarray}\label{energiapotencialdef}
	U(r_{+}) - U(r) = \int_{r_{+}}^{r}F_{n} dr,
	\end{eqnarray}
	
	\begin{eqnarray}\label{energiatencialdefintegracionfinal}
	U(r) =  -\frac{3 M c^2}{8}\Big[\Big(\frac{r}{r_{+}}\Big)^4 - 8 \Big(\frac{r}{r_{+}}\Big) + 7 \Big] +U(r_{+}).
	\end{eqnarray}
	
	To determine the potential $U(r)$ at a point $r$ inside the black hole, we need to know the potential energy on the surface of the black hole, $U(r_{+})$. $U(r)$ is the potential energy of  interaction between the black hole and the vacuum.
	
	\section{Qualitative Analysis of the Movement}
	
	We now consider the potential given by Eq. (\ref{energiatencialdefintegracionfinal})  and will try to describe qualitatively the motion of a classical particle of mass $m$ in this potential field, for the case when its energy is zero and $U(r_{+} ) = 0$. In this case we do not consider the angular momentum of the particle.
	
	\begin{eqnarray}\label{energiatencialdefintegracionfinal1}
	U_{bhv}(r) = - \frac{3 M c^2}{8}\Big[\Big(\frac{r}{r_{+}}\Big)^4 - 8 \Big(\frac{r}{r_{+}}\Big) + 7 \Big].
	\end{eqnarray}
	
	The area $1 < x < 1.5 $ is classically forbidden for a particle, considering its energy $\tilde{E}$ equal to zero. For photons, the surface of the event horizon is allowed, in which case $U(x) = 0$,  $x = 1$ and $x \approx 1.5 $, the two spheres of the photon. If the particle has energy $\tilde{E} < \tilde{U}_{max} = 0.2098$, it cannot escape the attraction of the black hole. The repulsive force on the right side of $x = \sqrt [3]{2}$ is stronger than  on the left side. If the particle has energy $\tilde{E} < 0$ and comes from infinity, the hole exerts a strong repulsive force on it. The opposite occurs when the particle comes from the left with energy $\tilde{E} < 0$. For the case when $\tilde{E} > \tilde{U}_{max}$ and the particle goes to the left, its position tends to infinity. For $\tilde{E} = 0.2098$, any perturbation tends to make the particle go to infinity or fall inside the black hole.
	
	\begin{figure}[!htb]
		\centering
		\includegraphics[width=0.6\linewidth, height=0.3\textheight]{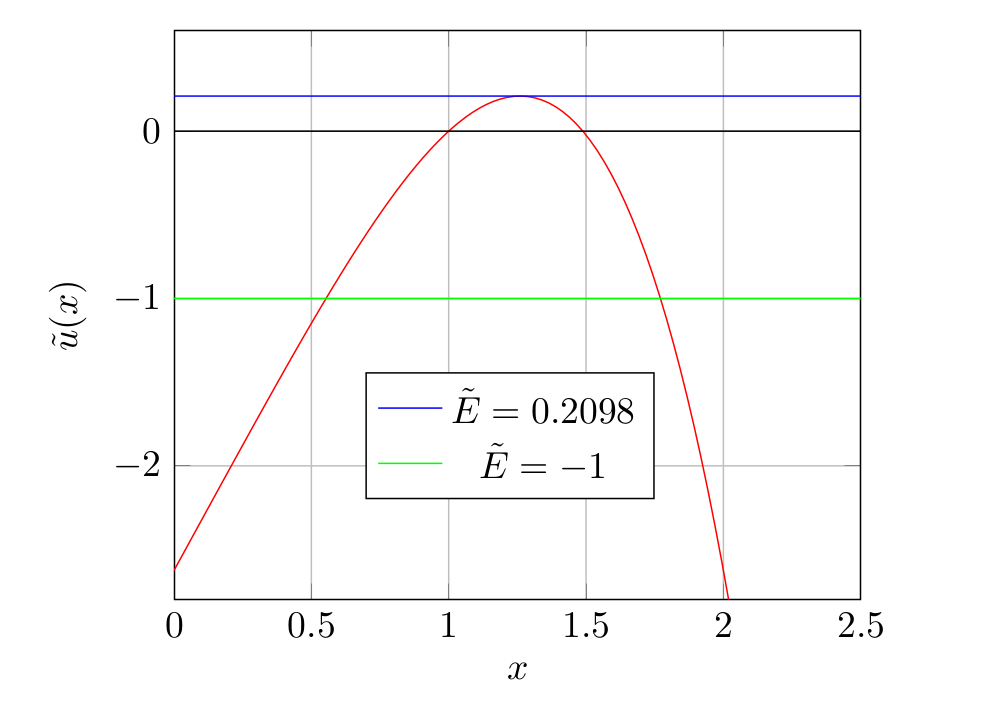}
		\caption{$\tilde{u}(x)$ is not symmetric with respect to $x = \sqrt[3]{2}$. Here $\tilde{u}(x) = U(x)/M c^2 $.}
		\label{FIGURA3}
	\end{figure}

	\section{The Hawking entropy}
	
	We can now determine the Hawking entropy in the same way as it is determined in thermodynamics, through its relationship with the amount of heat $\delta Q$,
	
	\begin{eqnarray}
	\Delta S = \int \frac{\delta Q}{T}.
	\end{eqnarray}
	
	We can imagine a process through which we add a certain amount of mass $dm$ to the black hole, which leads to the increase of its radius. For any radius of the event horizon we can write  Eq. (\ref{temperatureenriquefrontera}) as, 
	
	\begin{eqnarray}\label{temperatureenriqueeneral}
	T_{H} = \frac{\hbar c}{4 \pi k_{B} r}.
	\end{eqnarray}
	
	On the other hand we have that $\delta Q = dU$ ,  and considering Eq. (\ref{Energiainternablackhole}) we have,
	
	\begin{eqnarray}
	\Delta S = \frac{\pi k_{B} c^3}{\hbar G}\left[r_{+} - \left(2 l_{P}\right)^2\right].
	\end{eqnarray}
	
	If we despise the second term for its smallness, we get the desired result:
	
	\begin{eqnarray}
	\Delta S = \frac{ k_{B} c^3}{4 \hbar G} \mathcal{A}.
	\end{eqnarray}
	
	In this way we show a possible origin of the Hawking entropy.
	
	\section{Potential and effective energy density}
	
	Now we will determine the effective potential. Eq. (\ref{energiatencialdefintegracionfinal1}) refers to the potential energy of interaction between the black hole and the vacuum, thus, the potential due to the vacuum, $V_{vac}(r) = U(r)/M$, is given by
	
	\begin{align}\label{potencialvacio}
	\alignedbox{}{V_{vac}(r) = - \frac{3 c^2}{8}\Big[\Big(\frac{r}{r_{+}}\Big)^4 - 8 \Big(\frac{r}{r_{+}}\Big) + 7 \Big].}
	\end{align}
	
	\begin{figure}[h]
		\centering
		\includegraphics[width=0.6\linewidth, height=0.3\textheight]{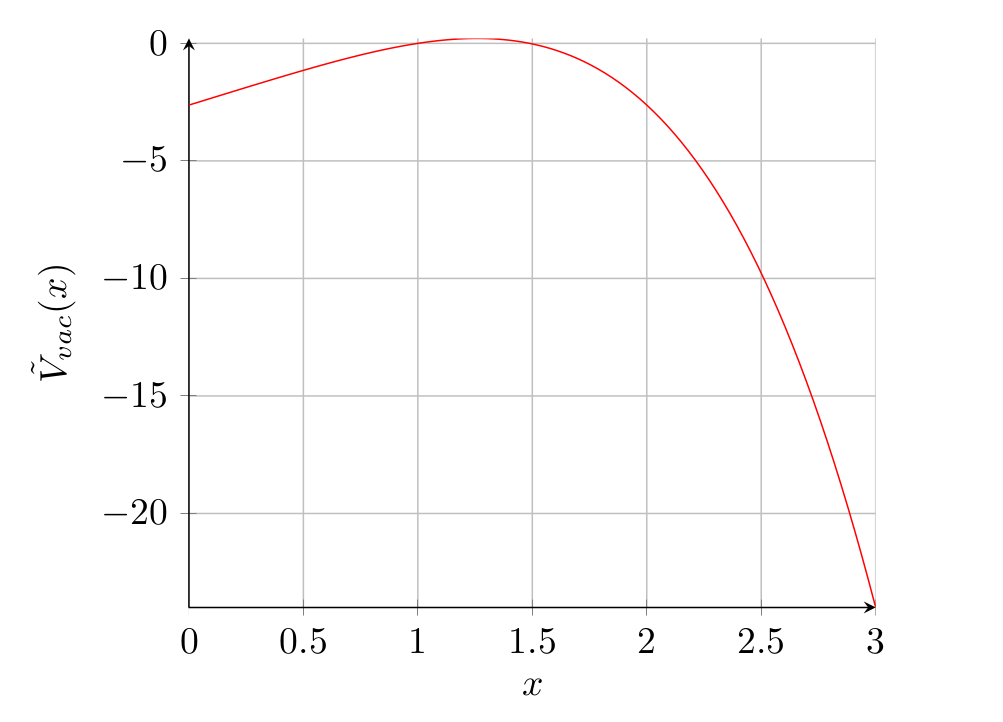}
		\caption{Potential of vacuum, where $\tilde{V}_{vac} = V_{vac}/c^2$, and $x = r/r_{+}$.}
		\label{FIGURA4}
	\end{figure}

	This potential is shown in Fig.\ref{FIGURA4}
	
	At any point in space, the total potential $V(r)$ (effective potential,$V_{eff}$) is given by the sum of the potential due to the vacuum, $ V_{vac}(r)$, and the potential due to the black hole, $V_{bh}(r)$. To find the potential due to the black hole we will consider the density given by Eq. (\ref{densityfinal}) and considering radial symmetry, we can solve the Poisson equation
	
	\begin{eqnarray}
	\nabla^{2}V_{bh} = -4\pi G \rho_{bh},
	\end{eqnarray}
	
	where
	
	\begin{eqnarray}
	\nabla^2 = \frac{2}{r} \frac{\partial}{\partial r} +  \frac{\partial^2}{\partial r^2},\nonumber 
	\end{eqnarray}
	
	then, solving for $V_{bh}$ we have
	
	\begin{eqnarray}\label{blackholepotential1}
	V_{bh}(r) = \frac{c_{1}}{r} -c_{2} - \frac{c^2}{2}\ln r,
	\end{eqnarray}
	
	To find the constants $c_{1}$ and $c_{2}$ we consider that $V_{bh}(r_{+}) = 0$ and $\nabla V_{bh}(r)\vert_{r = r_{+}} = \kappa$,  then we get
	
	\begin{align}\label{blackholepotential2}
	\alignedbox{}{V_{bh}(r) =  c^2(1 - \frac{r_{+}}{r}) + \frac{c^2}{2}\ln \frac{r_{+}}{r},}
	\end{align}
	
	\begin{figure}[h]
		\centering
		\includegraphics[width=0.6\linewidth, height=0.3\textheight]{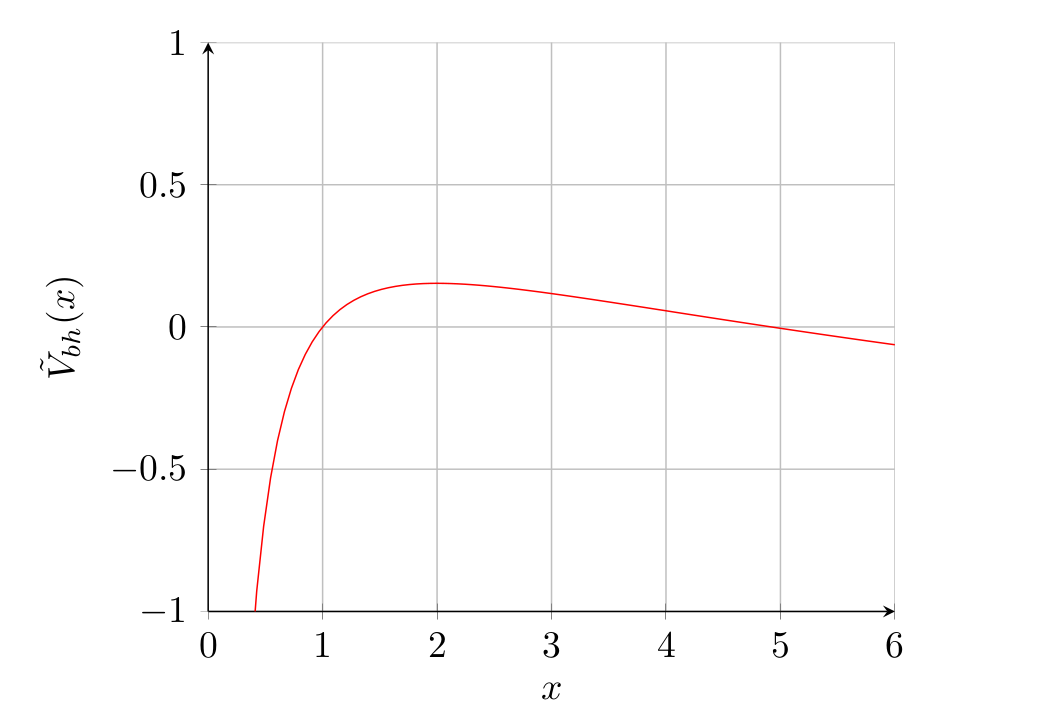}
		\caption{Black hole potential. It looks similar to the potential generated by the gravitational field of a body of mass M.}
		\label{FIGURA5}
	\end{figure}

	and the effective potential,
	
	\begin{align}\label{potenciaeffectivo}
	\alignedbox{}{V_{eff}(r) = -\frac{3 c^2}{8}\Big[\Big(\frac{r}{r_{+}}\Big)^4 - 8 \Big(\frac{r}{r_{+}}\Big) + 7\Big] +  c^2\Big(1 - \frac{r_{+}}{r}\Big) + \frac{c^2}{2} \ln \frac{r_{+}}{r}.}
	\end{align}
	
\begin{figure}[h]
		\centering
		\includegraphics[width=0.6\linewidth, height=0.3\textheight]{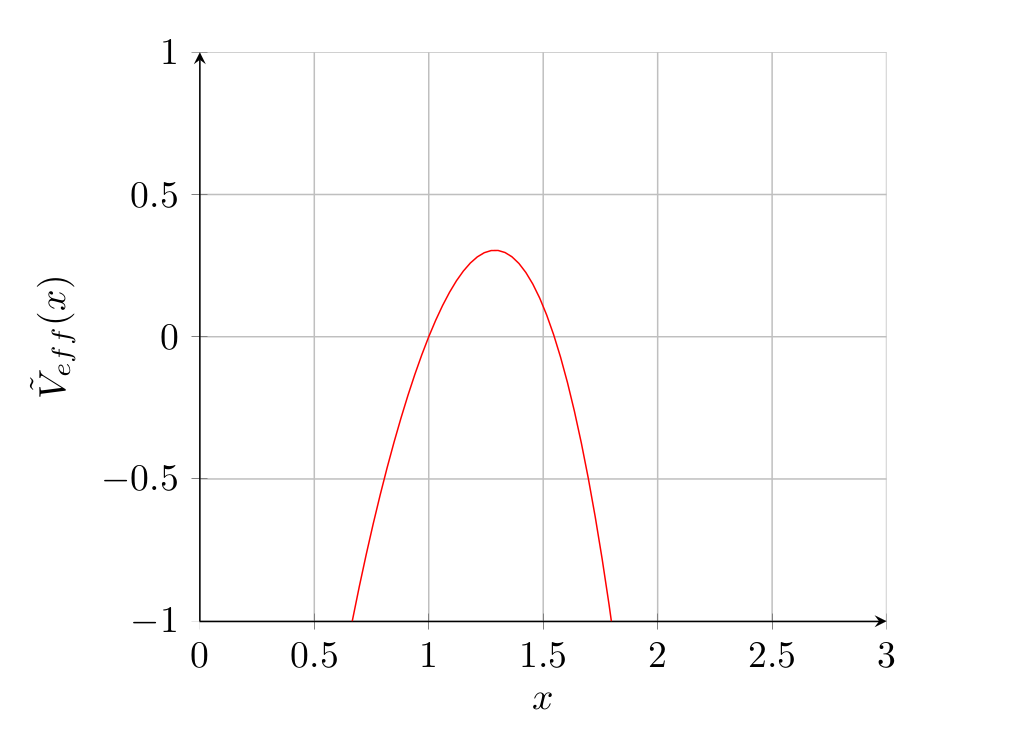}
		\caption{Effective potential. Shows the sharp drop due to the contribution of the vacuum potential.}
		\label{FIGURA6}
	\end{figure}

	Fig. \ref{FIGURA6} show this effective potential.
	
	\newpage
	
	The vacuum density can be found from the Poisson's equation, 
	
	\begin{eqnarray}
	\nabla^{2}V_{vac} = +4\pi G \rho_{vac},
	\end{eqnarray}
	
	considering spherical symmetry and the Eq.(\ref{potencialvacio}) , then we get that $\rho_{vac}$ is given by,

	\begin{align}\label{densidadvacio}
	\alignedbox{}{\rho_{vac} =-\rho_{+}\big[ 15\big(\frac{r}{r_{+}}\big)^2 - 12 \big(\frac{r_{+}}{r}\big)\big]}.
	\end{align}

	Eq.(\ref{densidadvacio})  offers us the relationship between the energy density of baryonic matter and the energy density of the vacuum inside the black hole, showing that both densities are of the same order, that confirm  the cosmic coincidence as refered \cite{zlatev1999quintessence}, and at the boundary, 
	
	\begin{align}\label{densidadenfrontera}
	\alignedbox{}{\rho_{vac} = -3\rho_{+}.}
	\end{align}
	

	On the other hand, Eq. (\ref{densidadvacio}) shows the existence of a surface inside the black hole in which the energy density of the vacuum is zero. This surface divides the interior of the black hole into two fundamental regions, one containing the attractive contribution of the energy density of the vacuum and the other containing the repulsive contribution, within the limits of the baryonic matter.
	
	\begin{figure}[h]
		\centering
		\includegraphics[width=0.6\linewidth, height=0.3\textheight]{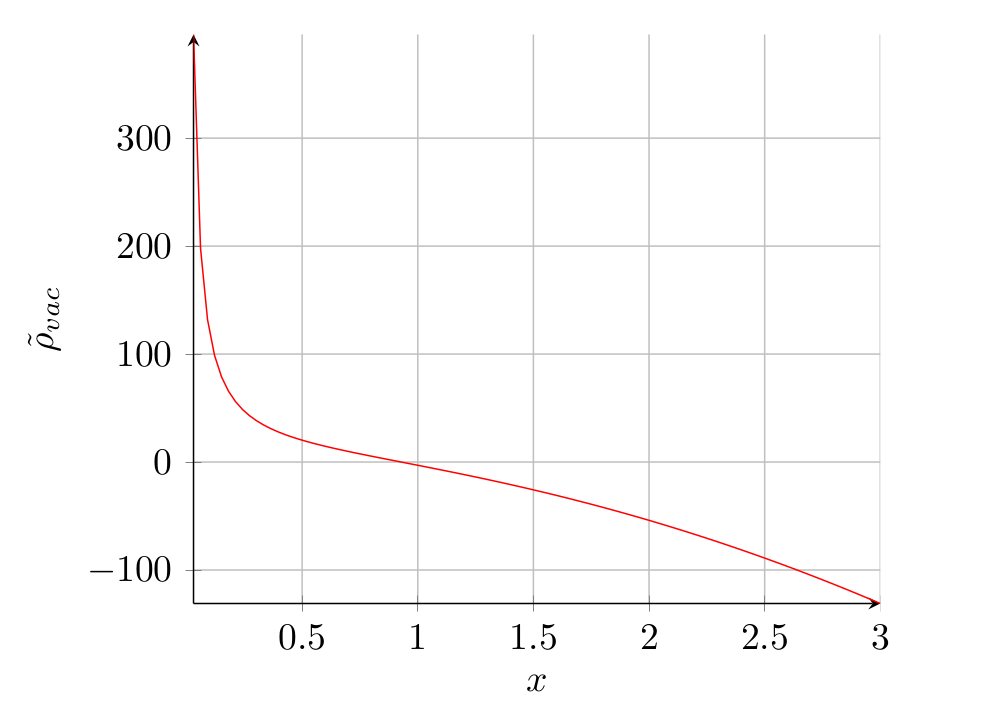}
		\caption{Vacuum density, showing a possitive region for $x < 1$, here $\tilde{\rho}_{vac} = \rho_{vac}/\rho_{+}$. }
		\label{FIGURA7}
	\end{figure}
	\hskip 10pt 
	\begin{figure}[h]
		\centering
		\includegraphics[width=0.6\linewidth, height=0.3\textheight]{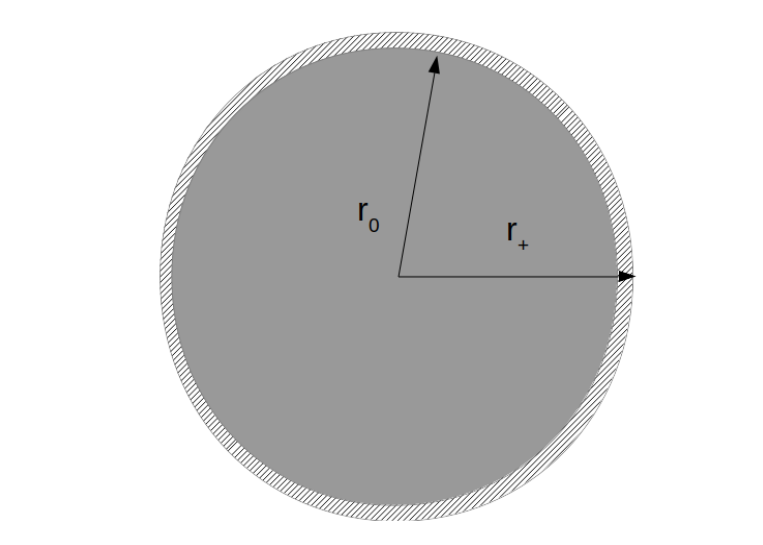}
		\caption{The region $2 l_{P} < r \leq r_{0}$ is attractive and $r_{0} <r \leq r_{+}$ has boths, attractive and repulsive parts.}
		\label{FIGURA8}
	\end{figure}

	It may be that this relationship is not exclusive to a black hole but rather a general law of nature. If these densities reffers to the same volume, then $m_{vac} = -3 M$. Eq. (\ref{densidadenfrontera}) shows the close link between matter and space and a dominance of vacuum energy .
	
	So the effect of the vacuum is exactly three times as large in its anti-gravity properties as the gravitational attraction of ordinary matter, which is actually sppeeding up because of this repulsion. Eq. (\ref{densidadvacio}) also shows us that for a certain value of $r = r_{0}$, the energy density of the vacuum, $\rho_{vac}$, is zero. This point ($r_{0} = \sqrt[3]{4/5} r_{+}$) is an inflection point. Note that to the left of that point the energy densities of the vacuum and baronic matter(given by Eq.(\ref{densityfinal})) are concave and to the right of that point,  the energy density of the vacuum is convex. Because  $\rho^{\prime\prime}_{vac}(r_{0} )= 0$, $r_{0}$ is a critical point of the second kind.
	
	From Eq.(\ref{densidadenfrontera}) we can see that $\rho_{vac}$ represent $75\%$ of the total energy density and $\rho_{+}$, $25\%,$ as we hope. The ratio between the attractive energy densities from baryonic matter and the energy density from a vacuum will be determined below.
	
	We hope that the vacuum density energy be negative in all space, but the Fig.\ref{FIGURA7} show us that for certain values of $x < 1$ it is positive, where $x = 1$ represent the surface of the  black hole. This fact has a relevant physical meaning, \textit{the corresponding possitive part of vacuum density energy is attractive}, maybe a new physical effect. This is big surprise because until now we belived that the vacuum energy density was repulsive. This makes us think and relate this mass obtained from the positive part of the energy density of a vacuum with dark matter. It seems that the extreme confinement of the energy of the vacuum or its interaction with baryonic matter makes it acquire attractive properties. On the other hand, it must be taken into consideration that if there is matter in a region then there is energy and where there is energy, there is potential and where there is potential, there is the possibility of any matter appearing or manifesting one of its common properties.
	
	Then, the effective energy density is given by
	
	\begin{align}\label{densidadefectiva}
	\alignedbox{}{\rho_{eff} = \rho_{+}\big[\big(\frac{r_{+}}{r}\big)^2- 15\big(\frac{r}{r_{+}}\big)^2 + 12 \big(\frac{r_{+}}{r}\big)\big],}
	\end{align}

\begin{figure}[!htb]
		\centering
		\includegraphics[width=0.6\linewidth, height=0.3\textheight]{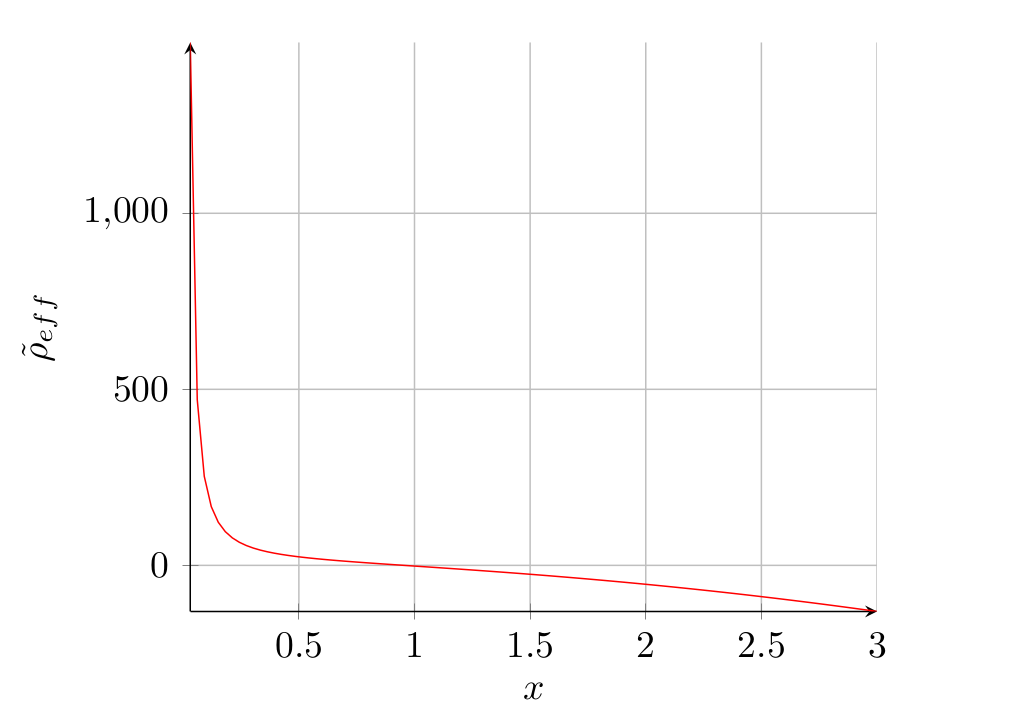}
		\caption{The Fig. shows the effective energy density and its positive region that comes from the contribution of the energy density of the baryonic matter object and that from a part of the energy density of the vacuum.}
		\label{FIGURA9}
	\end{figure}

	Fiq.\ref{FIGURA9} shows this effective density.
	
	Eq. (\ref{densidadefectiva}) show that for certain $r_{0} $, $\rho_{vac} = 0$. The corresponding actractive effective mass part, due to the positive effective energy density, could be found as, 
	
	\begin{eqnarray}
	m_{eff}^{+} &=& \int_{0}^{r_{0}}\rho_{eff} dv\\
	&=& \rho_{+}\int_{0}^{r_{0}}\Big[ \big(\frac{r_{+}}{r}\big)^2-\big[ 15\big(\frac{r}{r_{+}}\big)^2 - 12 \big(\frac{r_{+}}{r}\big)\big]\Big]4\pi r^2 dr\nonumber\\
	m_{eff}^{+} &\approx& 4.0441 M\nonumber
	\end{eqnarray}
	
	
	\begin{align}
	\alignedbox{}{m_{eff}^{+} \approx M + 3.0441 M},
	\end{align}
	
	the first term corresponding to the black hole mass density(baryonic part) and the second, to the vacuum energy density(may be the corresponding dark matter part), which percentage are $6$ and $19\%$  respectively. 
	
	On the other hand, repulsitive mass part of the effective energy density is
	
	\begin{eqnarray}
	m_{eff}^{-} &=& \int_{r_{0}}^{r_{+}}\rho_{eff} dv\\
	&=& -0.0441 M.\nonumber
	\end{eqnarray}
	
	Resulting that the total effective mass would be equal to
	
	\begin{eqnarray}\label{masaefectivatotalatractiva}
	m_{eff}^{Tot} &\approx& 4 M.
	\end{eqnarray}
	
	This excess mass ($3 M$)  roughly matches with the lack of mass reported in the analysis of galaxies rotating around galactic clusters. 
	
	\section{Field intensity}
	
	It is time to ask ourselves what the value of surface gravity would be if the effective mass is 4 times the mass of the black hole. Well, we can now calculate the intensities of the fields corresponding to the potentials according to the equation,
	
	\begin{eqnarray}
	\textbf{g} = \pm \nabla V,
	\end{eqnarray}
	
	plus sign for repulsive potential and minus for attractive potential.
	
	For the potential of the vacuum Eq. (\ref{potencialvacio}) we have to
	
	\begin{eqnarray}\label{gravedadvacio}
	\textbf{g}_{vac} = \frac{3 c^2}{r_{+}}\big[1 - \frac{1}{2}\big(\frac{r}{r_{+}}\big)^3\big]\textbf{r}.
	\end{eqnarray}
	
	whose value in the event horizon is given by
	
	\begin{eqnarray}\label{gravedadvaciohorizontedeeventos}
	\kappa_{vac} = \frac{3 c^4}{4 G M},
	\end{eqnarray}
	
	for potential Eq.(\ref{blackholepotential2}) we have
	
	\begin{eqnarray}\label{gravedadagujeronegro}
	\textbf{g}_{bh} = \frac{c^2}{r}\big[\frac{r_{+}}{r}-\frac{1}{2}\big]\textbf{r},
	\end{eqnarray}
	
	whose value for $ r = 2 r _ {+} $ and for $ r \to \infty $ is equals to zero and coincides with the value for surface gravity in $ r = r_ {+} $. Eq. (\ref{gravedadagujeronegro}) is very similar to that from \cite{bardeen1973four}. The maximum theoretical value from the classical point of view when $r = \mathit{l}_{P}$, is given by
	
	\begin{equation}\label{surfacegravity3}
	\kappa = \frac{c^2}{2 \mathit{l}_{P}}\big[\frac{4 M}{m_{P}} - 1\big].
	\end{equation}
	
	The flows of vectors Eqs. (\ref{gravedadvacio}) and (\ref{gravedadagujeronegro}) evaluated in the event horizon obey Gauss's law for these fields, giving  $-4\pi M$ and $-4\pi(3M)$ respectively, as expected.
	
	Then, considering these two effects the effective surface gravity will be
	
	\begin{eqnarray}\label{gravedadsuperficialefectiva}
	\textbf{g}_{eff} = \frac{c^4}{ M G} \textbf{r}.
	\end{eqnarray}
	
	This is an expected result and is due to the attractive contribution exerted by the energy density of the vacuum inside the black hole. The Eq. (\ref{gravedadsuperficialefectiva}) shows that the event horizon has a gravity 4 times greater than when the mass of the vacuum is not taken into account.
	
	\section{The model and its relationship with some characteristics of our universe}
	
	There are innumerable physical models developed in a specific area to explain certain phenomena which find applications in similar or different areas, perhaps it is the case of the model developed in this work. As we will show, there is an apparent relationship between the proposed model and some fundamental characteristics of our universe.
	
	The model tries to characterize the interior of a black hole including its surface and on this we will focus the following analysis, using various observational astronomical data.
	
	\subsection{Cosmological constant and density of the vacuum}
	
	In the previous section we determined the energy density of the vacuum. This allows us to establish a link between this physical magnitude and the cosmological constant $\Lambda$, considering the definition of the latter,
	
	\begin{eqnarray}
	\rho_{vac} \equiv -\frac{c^2}{8 \pi G}\Lambda.
	\end{eqnarray}
	
	Taking into account Eq. (\ref{densidadvacio}) we see that

	\begin{align}
	\alignedbox{}{\Lambda = \frac{1}{r_{+}^2}\big[15\big(\frac{r}{r_{+}}\big)^2 -12\big(\frac{r_{+}}{r}\big)\big],}
	\end{align}\label{lambda}
	
	it is not a constant and can have positive, negative, or null values. Let's analyze the case of $r = r_{+}$, for which,
	
	\begin{eqnarray}
	\Lambda = \frac{3}{r_{+}^2},
	\end{eqnarray}\label{lambdafrontera}
	
	this result matches with \cite{silverstein2017tasi}. Now we are going to consider $r_{+} = 4.4 \times 10^{26}\, \unit{m} $ (radius of the observable universe), we get, 
	
	\begin{eqnarray}
	\Lambda = 1.5496\times 10^{-53}\,\unit{m}^{-2},
	\end{eqnarray}\label{lambdauniverso}
	
	value that coincides with that mentioned in the literature.
	
	\begin{figure}[h]
		\centering
		\includegraphics[width=0.6\linewidth, height=0.3\textheight]{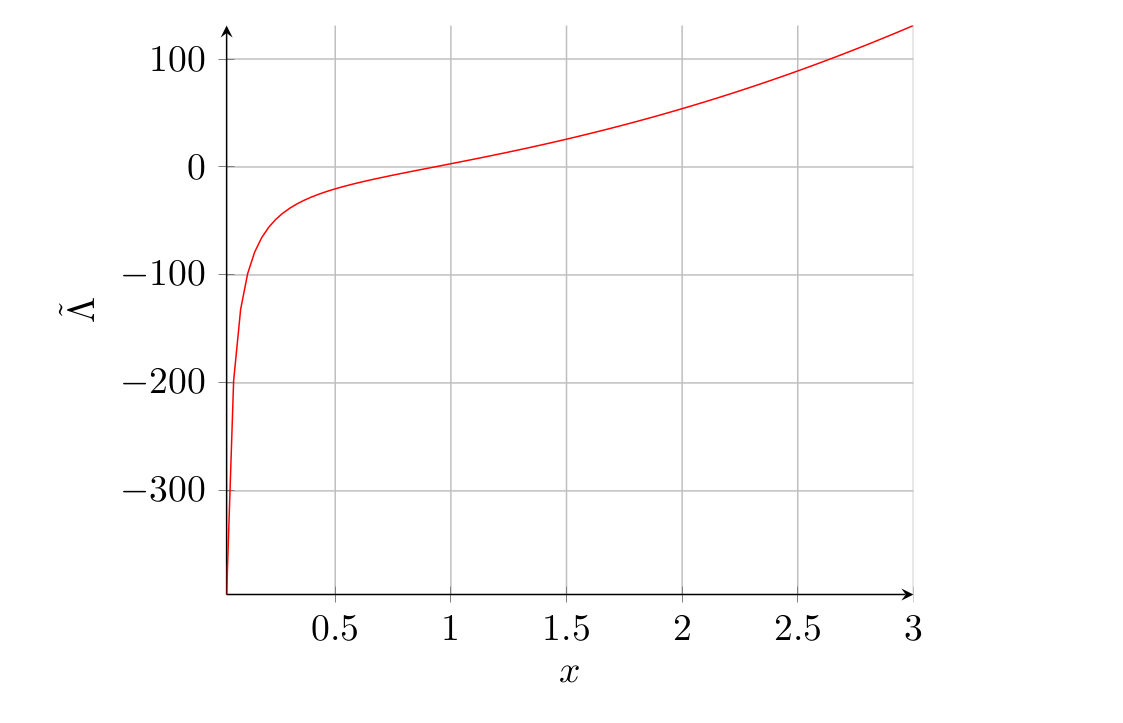}
		\caption{For values less than $r_{0}$ lambda is negative, the zone corresponding to the interior of the mass distribution. For values greater than $r_{0}$ it is positive, and null for $r = r_{0}$. Here $\tilde{\Lambda} = \Lambda r_{+}^{2}$.}
		\label{FIGURA10}
	\end{figure}

	In the case of the vacuum density our model can estimate this value considering the mass being the mass of the universe, $m_{u} \approx 1.50\times 10^{53}\,\unit{Kg}$ to get,
	
	\begin{eqnarray}
	\rho_{vac} &=& -\frac{3 c^6}{32 \pi G^3 m_{u}^2}\\
	&\approx& -3.23\times 10^{-30}\,\unit{g/cm^3},\nonumber 
	\end{eqnarray}
	
	which is less than the critical density $\rho_{c} = 8.40 \times 10^{-30}\,\unit{g/cm^3}$, and coincide with that $\vert \Omega_{vac}\vert = \vert \rho_{vac} / \rho_{c} \vert = 0.38 \leq 1$.
	
	For  $m_{u} \approx 1.00 \times 10^{53}\,\unit{Kg}$, we have that $\rho_{vac} = -7.29\times 10^{-30}\,\unit{g/cm^{3}}$, $\vert \Omega_{\Lambda}\vert \approx 0.87$, $\vert \Omega_{c}\vert \approx 0.22$  and  $\vert \Omega_{b}\vert \approx 0.06$, then $\vert \Omega_{T}\vert \approx 1.16$.
	
	\subsection{Hubble constant}
	
	Now we are going to consider the Hubble constant. Hubble's Law is often expressed by the equation $v = H_{0} d$, where $H_{0}$ is the proportionality constant (Hubble's constant) between the distance $d$ for a galaxy and its velocity $v$,
	
	\begin{eqnarray}
	H_{vac} \equiv \sqrt{\frac{8 \pi}{3}G\vert \rho_{vac}\vert},
	\end{eqnarray}
	
	taking into accont Eq. (\ref{densidadvacio}) we have that

	\begin{align}
	\alignedbox{} {H_{vac} = \frac{c}{r_{+}}\sqrt{15 x^2 - 12 x^{-1}},}
	\end{align}
	
	\begin{figure}[!htb]
		\centering
		
		\label{CONSTANTEHUBBLE}
		\subfloat[On close scales, the "constant" shows an apparent linearity with respect to $x$. \label{HUBLEPEQUEÑADIST}]{\includegraphics[width=0.50\linewidth, height=0.3\textheight]{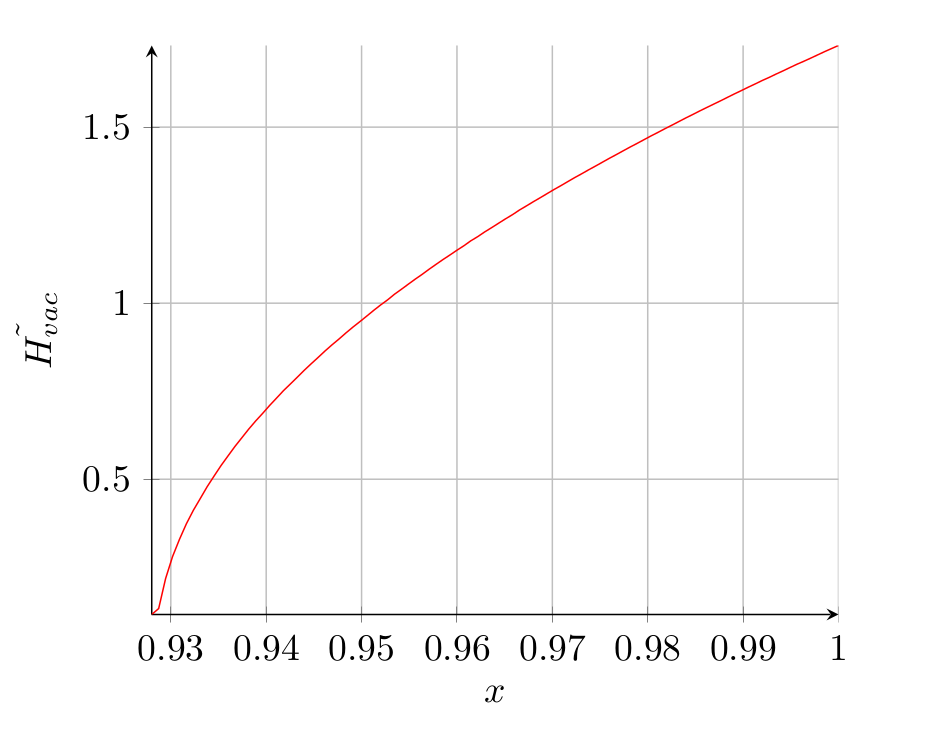}}
		\hfill
		\subfloat[At large scales the "constant" shows a strong tendency towards linearity with respect to $x$.\label{HUBLEGRANDEDIST}]{\includegraphics[width=0.50\linewidth, height=0.3\textheight]{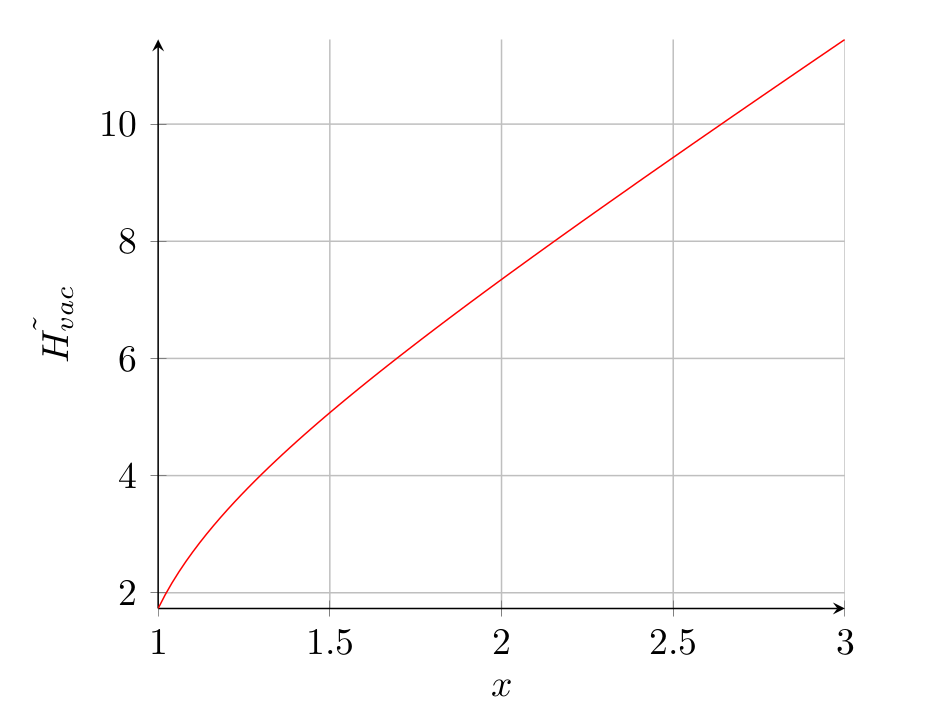}}
		\caption{The curves show that $H_{vac}$ is not constant after all. The curve on the right side is for illustrative purposes.}
	\end{figure}

	and for $r = r_{+}$, we have that,
	\begin{eqnarray}
	H_{vac} = \frac{c\sqrt{3}}{r_{+}}.
	\end{eqnarray}\label{constantehubbleenlafrontera}
	
	Considering $r_{+} = 2 m G/c^2$, and using the mass of universe $m = m_{u} = 1.5 \times 10^{53} \,\unit{Kg}$ we have,
	
	\begin{eqnarray}
	H_{vac} &=& 2.33 \times 10^{-18} \, \unit{s^{-1}}\\
	&=& 71.86 \, \unit{K m/s/Mpc},\nonumber
	\end{eqnarray}\label{constantehubbleenlafronteravalor}
	
	value very close to that reported in \cite{bennett2011seven} and \cite{hinshaw2013nine}.
	
	As we can see, the Hubble constant is not constant after all, which has an implication in the expansion speed of galaxies and Hubble's law, which will not be a linear speed, but quadratic on large scales.

	\begin{align}
	\alignedbox{}{v = c x\sqrt{\vert15 x^2 - 12 x^{-1}\vert}}
	\end{align}\label{leydehubble}
	
	\begin{figure}[!htb]
		\centering
		\subfloat[The velocity shows an apparent linearity with respect to $x$.\label{FIGURA12A}]{\includegraphics[width=0.5\linewidth, height=0.3\textheight,]{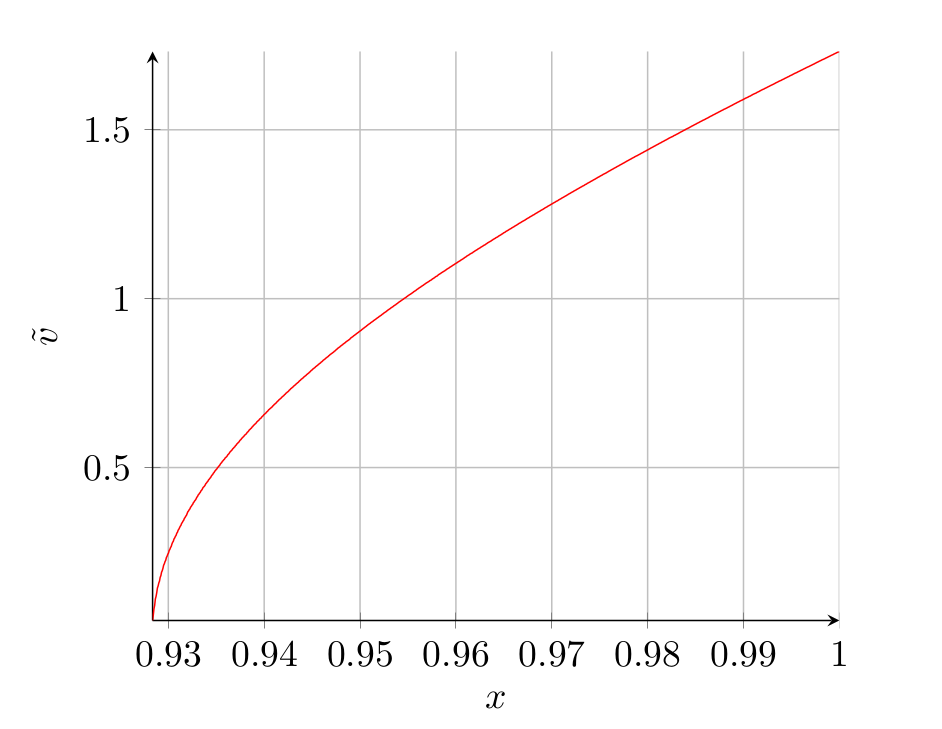}}
		\hfill
		\subfloat[On large scales the speed shows its slight parabolic tendency.\label{FIGURA12B}]{\includegraphics[width=0.5\linewidth, height=0.3\textheight]{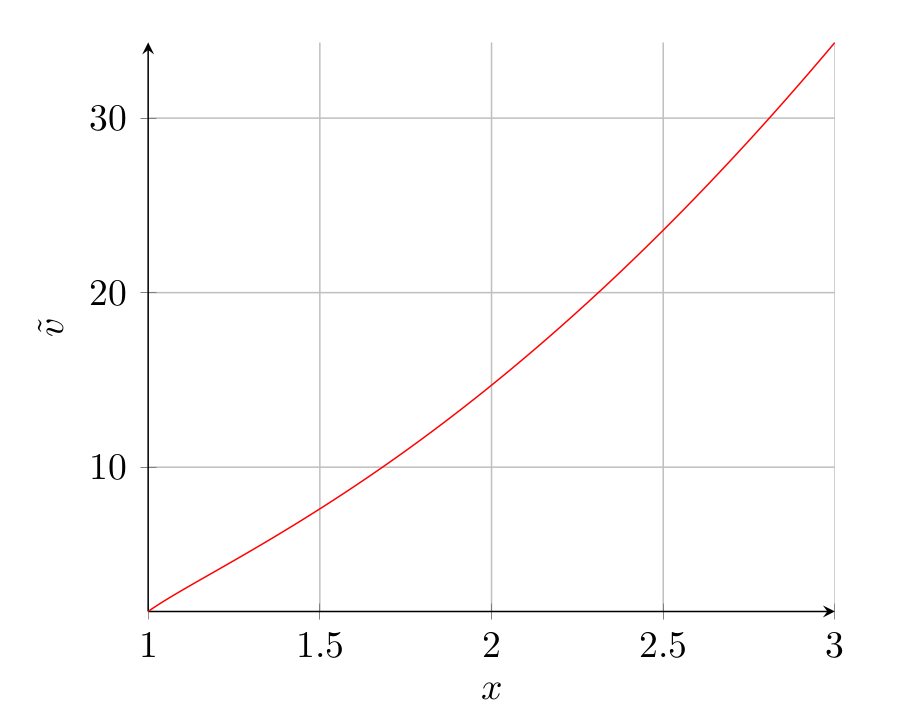}}
		\caption{The graphs show the apparent linearity of Hubble's law. The figure on the right side is for illustrative purposes.}
		
	\end{figure}

	\subsection{Galaxy rotation curve}
	
	The experimental curve of rotation of the galaxies does not fit the rotation curve corresponding to the theoretical model. The mass of the cluster to which that galaxy belongs is not enough to keep the galaxy rotating around its center. One explanation for this is the existence of dark matter in that region, with sufficient mass that would allow the galaxies to rotate with the observed speed.Then, there is a significant discrepancy  between the experimental curves observed, and the obtained by applying gravity theory to the matter observed in a galaxy. Here we will tries to explain this problem involving the effective mass $m_{eff}^{+}$,  and to stablish a link between this mass and the so called \textit{dark matter}. The problem is solved by considering an appropiated mass function distribution $m_{eff}^{+}(r)$.
	
	The effective mass $m_{eff}^{+}$ as function of $r$ is given by,
	
	\begin{align}\label{masaefectivavacio}
	\alignedbox{}{m_{eff}^{+}(r) = \frac{c^2 r}{2 G}\big[1 -3 (\frac{r}{r_{+}})^4 + 6 (\frac{r}{r_{+}}) \big].}
	\end{align}
	
	Now we can use directly the know escape velocity,
	
	\begin{eqnarray}\label{velocidaddeescape1}
	v = \sqrt{\frac{2 m_{eff} G}{r}},
	\end{eqnarray}
	
	but as we are considering relativistic effects we are going to use the relativistic kinetic energy to obtain $v$. Considering this, we get that

	\begin{align}\label{velocidaddeescaperelativistica}
	\alignedbox{}{v = c \sqrt{1 - \Big[\frac{1}{1 + \frac{G m_{eff}}{c^2 r}}\Big]^2}.}
	\end{align}
	
	Substituting Eq. (\ref{masaefectivavacio}) in (\ref{velocidaddeescaperelativistica}) we get,
	
	\begin{eqnarray}\label{velocidaddeescaperelativisticafinal}
	v = c \sqrt{1 - \Big[\frac{1}{\frac{3}{2} - \frac{3}{2}(\frac{r}{r_{+}})^4 + 3 (\frac{r}{r_{+}})}\Big]^2}, 
	\end{eqnarray}
	
	or
	
	\begin{eqnarray}\label{velocidaddeescaperelativisticafinalnormalizada}
	\tilde{v} =  \sqrt{1 - \Big[\frac{1}{\frac{3}{2} - \frac{3}{2}x^{-4} + 3 x^{-1}}\Big]^2}.
	\end{eqnarray}

	\begin{figure}[!htb]
		\centering
		\includegraphics[width=0.6\linewidth]{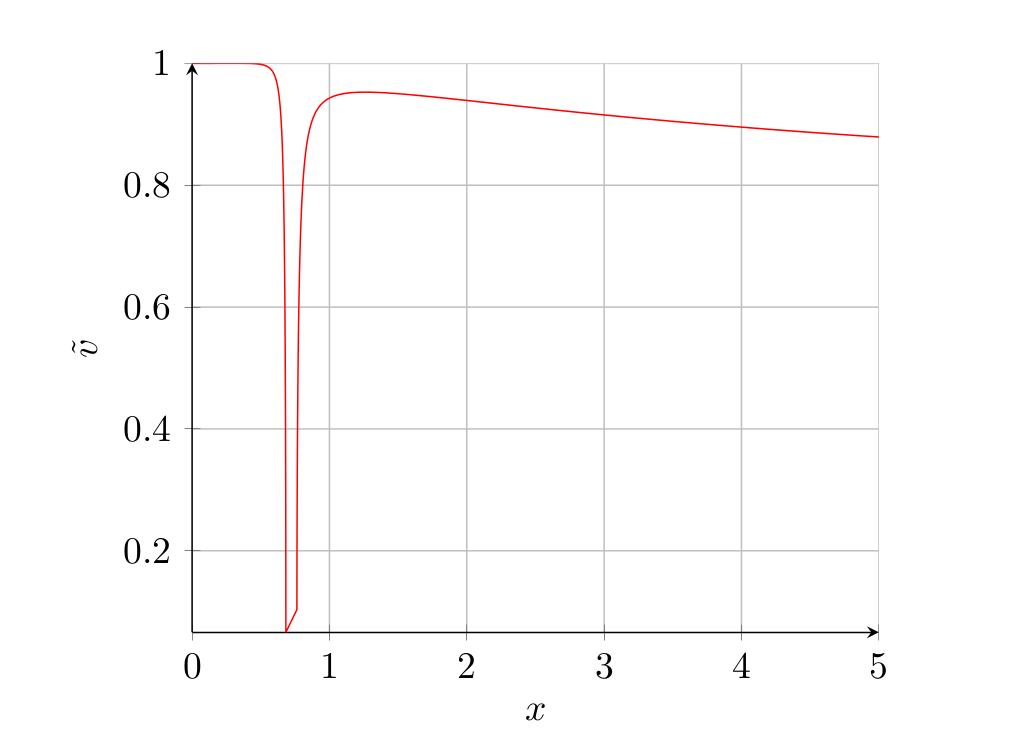}
		\caption{The curve show that do not decrease in the expected inverse square root relationship but is "flat". If $x \to \infty$, $v \to \sqrt{5}/3 c$.}
		\label{FIGURA13}
	\end{figure}

	As the galaxy mass increasse as $r$ and $r_{+}$ does, the key is plot $v$ as function of $r_{+}(x^{-1})$. This result is in agreement with the observational data. Fig. \ref{FIGURA13} shows a value for which the escape velocity is equal to zero. This result was expected and is due to the joint action of the attractive forces coming from the massive object and the positive contribution of the energy of the vacuum and the repulsive force due to the vacuum. At that point the effective acceleration equals zero.
	
	\section{Expanding universe}
	
	Why is the universe expanding? First of all, in our opinion the universe does not expand and the baryonic matter contained in it does, due to the difference between the energy densities of the vacuum and the baryonic matter. If we look again at our simplest model, which consists of a Schwarzschild  black hole and the vacuum, the net force $F_{n}$ shows us that as the radius $r_{+}$ increases, i.e. the mass increases, the black hole force, $F_{bh}$, decreases.
	
	\begin{eqnarray}\label{fuerzanetablackhole1}
	F_{n}(r) = \frac{3}{2}\frac{c^4}{G}\left[\frac{1}{2}\left(\frac{r}{r_{+}}\right)^3 - 1\right].
	\end{eqnarray}

	\begin{figure}[!htb]
		\centering
		\includegraphics[width=0.6\linewidth, height=0.3\textheight]{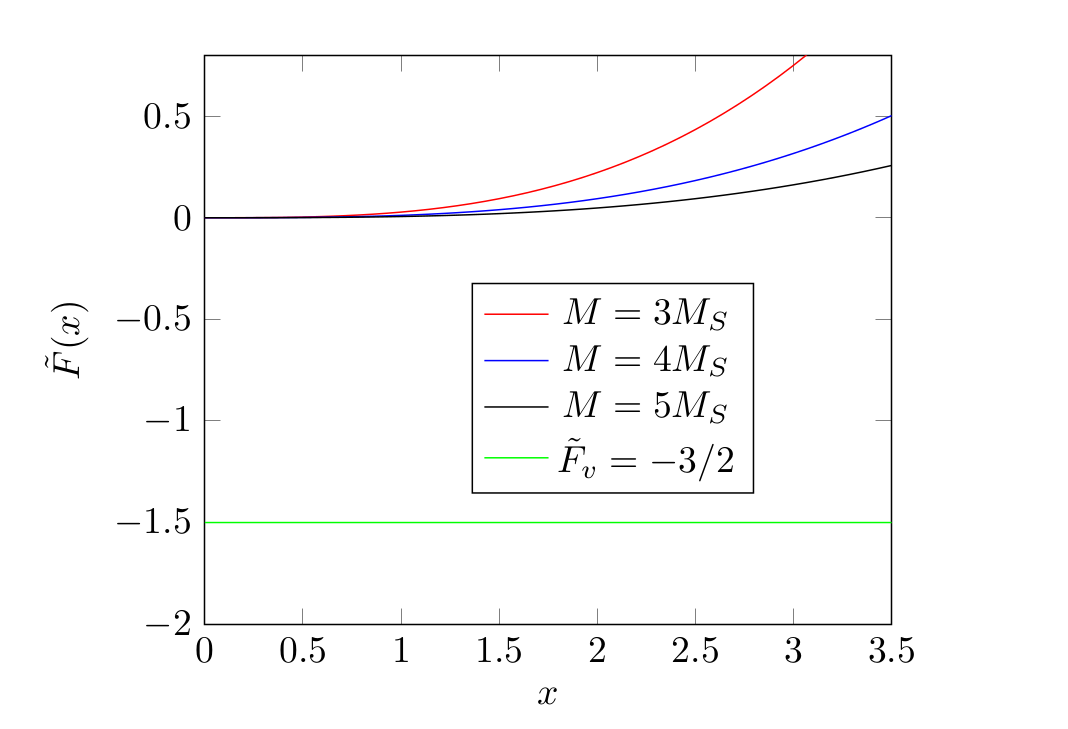}
		\caption{The black hole force $F_{bh}$ (for three values of the mass$M$: $3M_{S}$,  $4M_{S}$ and $5M_{S}$) and the vacuum force $\tilde{F}_{v}$, $\tilde{F}(x) =F_{n}(x)/ c^4/G$.} \label{fig:4}
		\label{FIGURA14}
	\end{figure}

	As $r_{+}$ increases, the black hole force $F_{bh}$ decreases, but the net force $F_{n}$ on any black hole surface increases. If we assume the early universe to be a black hole containing a finite amount of matter, then this force tends to expand it as the amount of matter increases. If the amount of matter increases by some process, the net force increases and approaches the maximum value $3/2\, c^4/G$ as the mass tends toward infinity. This is why the universe is expanding at an accelerated rate. The order of magnitude of this maximum force is $F_{n}(0)\sim 10^{44}\,\unit{N}$. From an energy point of view, Eq. (\ref{densidadvacio}) shows that the corresponding energy of matter with attractive property is greater than that of a vacuum(the part that is negative), this leads baryonic matter to tend to a state of lower energy infinitely expanding.

	Another problem is the order of magnitude of the energy density of a vacuum and baryonic matter when the universe expands.
	If we define, $\Delta \equiv \rho_{vac}/\rho_{+}$, then the limit when $r\to\infty$  and $r_{+} \to \infty$ is equals to,
	
	\begin{align}
	\alignedbox{}{\lim\limits_{\substack{r\to \infty\\  r_{+} \to \infty} }\Delta = -3,}
	\end{align}
	
	which indicates that even though the universe expands, the orders of these magnitudes remain the same.

	\section{Conclusions}
	
	The model assumes the existence of a type of black hole whose density is inversely proportional to the square of its radius and its temperature inversely proportional to its density.
	
	Treating the mass density of the black hole as a function of $\rho \propto 1/r^{2}$ and its temperature as $T \propto 1/\rho$, we investigate its intrinsic properties. We found for a given black hole that both the temperature and entropy have a parabolic dependence on the ratio $r$. Considering the black hole equation, we obtain its equations of state as a function of area $\mathcal{A}$. A conservative force and the corresponding potential energy were also obtained. This net conservative force is composed of two forces, one corresponding to the black hole and the other to the vacuum. We show that in the black hole limit the vacuum force is larger than that of the black hole and that at the event horizon the pressure is negative. We make a qualitative analysis of the motion in the potential field and show the existence of a forbidden region for the total energy equal to zero, from the event horizon to $1.5$ of this radius, both corresponding to the light spheres. We think that the Hawking entropy has its origin in the increase of the radius of the black hole due to the increase of its mass. We consider that dark matter is closely related to dark energy, being that dark matter is a state of dark energy that acquires attractive properties due to its interaction with baryonic matter, perhaps that is one reason why dark matter could not be detected, since it comes from dark energy.This result tells us that both Newton's laws and those of General Relativity are still valid even on large scales, so they do not have to be modified.The model shows that the composition of our universe is 75\% dark energy, 19\% dark energy and 6\% baryonic matter.  The model shows that dark matter is nothing more than a state of dark energy. The problem of the non-coincidence between the rotation curves of the galaxies, theoretical and experimental, is solved considering an appropriate density of matter distribution, which was offered by the model. We can briefly explain the accelerated universe, taking into account the model studied and the two forces acting on the black hole and we consider that the universe does not expand and the baryonic matter contained in it does. From an energy point of view, the corresponding energy of matter with attractive property is greater than that of a vacuum(the part that is negative), this leads baryonic matter to tend to a state of lower energy infinitely expanding. Finally, we consider that other physical theories may take these results into account to try to explain certain phenomena or anomalous behaviors observed.\\

	\bibliographystyle{acm}  
	
	\bibliography{references}
\end{document}